\documentclass{emulateapj}

\def\msun{{\rm ~M}_{\odot}}

\begin{document}

\title{The Contribution of Halo White Dwarf Binaries to the {\em LISA} Signal}

 \author{Ashley J. Ruiter,\altaffilmark{1,2} 
Krzysztof Belczynski,\altaffilmark{3,1} 
Matthew Benacquista,\altaffilmark{4} and Kelly
Holley-Bockelmann\altaffilmark{5}}

\email{aruiter@nmsu.edu, kbelczyn@nmsu.edu, benacquista@phys.utb.edu, k.holley@vanderbilt.edu}

 \altaffiltext{1}{New Mexico State University, Dept. of Astronomy,
        1320 Frenger Mall, Las Cruces, NM 88003}
 \altaffiltext{2}{Smithsonian Predoctoral Fellow, Center for
        Astrophysics, 60 Garden St., Cambridge, MA 02138}
 \altaffiltext{3}{Los Alamos National Laboratory, Los Alamos, NM, Oppenheimer Fellow}
 \altaffiltext{4}{Center for Gravitational Wave Astronomy, The University of Texas at 
Brownsville, 80 Fort Brown, Brownsville, TX 78520}
 \altaffiltext{5}{Department of Physics and Astronomy, Vanderbilt University, Nashville, TN 37235}

\begin{abstract} 
Galactic double white dwarfs were postulated as a source of confusion
limited noise for {\em LISA}, the future space-based 
gravitational wave observatory. Until very recently, the
Galactic population consisted of a relatively well studied disk
population, a somewhat studied smaller bulge population and
a mostly unknown, but potentially large halo population. It has been argued 
that the halo population may produce a signal that is much stronger (factor of 
$\sim 5$ in spectral amplitude) than the disk population. However, this 
surprising result was not based on an actual calculation of a halo white dwarf 
population but was derived on (i) the assumption that one can extrapolate the halo 
population properties from those of the disk population
and (ii) the postulated (unrealistically) high number
of white dwarfs in the halo.    
We perform the first calculation of a halo white dwarf population using
population synthesis models. Our comparison with the signal arising
from double white dwarfs in the Galactic disk$+$bulge clearly shows that it is
impossible for the double white dwarf halo signal to exceed that of
the rest of the Galaxy. 
Using microlensing results to give an upper limit on the content of white dwarfs in 
the halo ($\sim 30\%$ baryonic mass in white dwarfs), our predicted halo signal
is a factor of 10 {\em lower} than the disk$+$bulge signal.  Even in the implausible case
where all of the baryonic halo mass is found in white dwarfs, the halo signal does not 
become comparable to that of the disk$+$bulge, and thus would still have a negligible effect on 
the detection of other {\em LISA} sources. 
 \end{abstract}


\section{INTRODUCTION}

The Laser Interferometer Space Antenna ({\em LISA}) is a proposed joint
ESA/NASA mission that will be the first space-based gravitational
radiation (GR) detector \citep[see e.g.,][and references
therein]{Hug06}.  It has been known for some time \citep{Hil90} that
Galactic double white dwarfs will be a prominent source of GR
for {\em LISA}.  Thousands of Galactic double white dwarfs are expected to be
resolved well enough to yield their masses and orbital parameters
\citep{Nel01,Nel04,RBBL}, which will lead to an improved understanding of common
envelope evolution scenarios and the origin of Type Ia Supernova
and/or subdwarf B star progenitors \citep{Liv89,Web84,IT84,Han03}.
However, a much larger number of double white dwarfs ($\sim 10^{7}$) will be
detectable within the {\em LISA} sensitivity range but will be
unresolved.  In fact, close double white dwarfs are so numerous that they are
expected to dominate the {\em LISA} GR signal at low frequencies,\footnote{$f_{\rm
  gr}=2/P_{\rm orb}$ for circular binaries.} their
signal rising above that of the instrumental noise level and
generating confusion-limited noise - a confusion `foreground'.  Thus, in
order to attempt to uncover sources beneath the confusion noise, one
must construct the expected total GR signal from these binaries and remove it
from the {\em LISA} data stream. To do this, it is of considerable
importance to determine a priori the characteristics which will set the level
of signal (e.g., masses, orbital periods, location in the Galaxy, etc.) of the
double white dwarf population.  In addition, the level of signal arising from
Galactic double white dwarfs can be useful in constraining the
structure and extent of the Galactic thick disk \citep{BH04}.

The contribution of low- and intermediate-mass extragalactic binaries to
the {\em LISA} signal was shown to be rather insignificant in comparison with the 
Galactic populations \citep{FP03}. 
The GR signal from Galactic and extragalactic black hole  
MACHO\footnote{Massive Astrophysical Compact Halo Object, or sometimes
MAssive Compact Halo Object.} binaries was 
investigated by \citet{Iok99} and it was in general found that black hole binaries at 
cosmological distances will have a higher impact on {\em LISA} than will the halo black hole 
MACHOs, although the foreground signal from Galactic double white dwarfs will still dominate.  

The {\em LISA} GR signal arising from the Galactic population of double
white dwarfs has been investigated by several groups
\citep[e.g.,][]{Hil90,PP98,HB00,Nel01,Nel04,BDL04,Edl05,Tim06,RBBL}.
In most previous calculations, the Galactic GR signal was calculated
for a single component disk with the bulge excluded, until recently where 
a bulge component has been considered \citep{Nel04,RBBL}.  We note that
while the signal arising from the bulge should not be discounted, the
Galactic disk is the major contributor to the {\em LISA} GR signal out 
of the two populations (see \S~\ref{results} for further explanation).  
Up until now, a full calculation of the {\em LISA} GR signal due to the halo
population of binaries has never been calculated.  \citet{His00}
estimated that a Galactic halo white dwarf population will produce a GR signal
significantly exceeding that of a disk population, though it was
hypothesized by \citet{Nel01} that such a strong GR signal from the halo is
unlikely given the physical characteristics of a $\gtrsim 10$ Gyr
halo stellar population.  We demonstrate that the latter is indeed
true, and argue against the existence of a strong halo GR signal with our detailed
calculations \S\,4.

In this study, we calculate the {\em LISA} GR signal arising from white dwarf
binaries in the Galactic halo.  We include in our signal calculation any double 
white dwarf within the {\em LISA} sensitivity range: $f_{\rm gr} = 0.0001 - 0.1$ Hz
(orbital periods between $\sim 5.6$ hours and 20 s).  
We note here that this work is a followup study to our previous, more
detailed study of the {\em LISA} signal arising from double white
dwarfs in the Milky Way disk and bulge.  Thus, for a thorough 
description of population synthesis modeling, signal calculations
or binary evolutionary histories, we refer the reader to
that study \citep[][which we will refer to from now on as RBBLW]{RBBL}.
The main objective of this work is to determine whether or
not the halo population, omitted in most previous studies, will
provide a significant contribution to the {\em LISA} GR signal.  Since we do
not know the contribution of white dwarfs, in number or in mass, to the
Galactic halo, we use two extreme models to bracket our
uncertainties. In one model we set the halo contribution to zero
(e.g., no white dwarfs in the halo and the halo white dwarf baryonic mass fraction
$\eta_{\rm B}=0$), while in the other we assume that all of the
baryonic halo mass is in white dwarfs ($\eta_{\rm B}=1$).  We then calculate
models within these brackets to compare the various halo realizations
to that of the Galactic disk$+$bulge population to determine if (and at what point) the
GR signal from halo double white dwarfs becomes significant compared to that of
the rest of the Galaxy.  We discuss our results in context of recent microlensing
experiments in order to constrain our intermediate models, and comment
on the prospects for future {\em LISA} observations in \S~\ref{discussion}.
In \S~\ref{model} we describe our calculations, and in
\S~\ref{results} we present our results of the halo gravitational wave
signal juxtaposed with the Galactic (disk and bulge) gravitational foreground 
calculations of RBBLW and \citet{Nel04}.

\section{MODEL DESCRIPTION}
\label{model}

The Galactic baryonic halo mass, potential, and shape are all poorly constrained
\citep{zinn:1985:gcs, saha:1985:rls, morrison:1996:htm, majewski:1993:gss, morrison:2003:map}.
In general, halo properties are estimated from star counts of thousands of intrinsically 
bright stars, such as red giants, blue horizontal branch stars, and RR Lyrae variables, 
with associated photometric or spectroscopic parallaxes.
There is strong evidence that at least some of the halo was built by
destroying satellite galaxies, resulting in a ``stringy'' halo
structure~\citep{johnston:1999:tsp}. Some obvious evidence for this may be found in 
the Magellanic Stream, a tidal arc of stars
stretching over 100$^\circ$ of the sky~\citep{putman:1998:tdm, morrison:2000:mgh}.
Nonetheless, for this first attempt, we employ the canonical well-mixed
spherically symmetric halo with a total baryonic mass of $10^9$
M$_\odot$ and a density profile as follows
\citep{zinn:1985:gcs,morrison:1996:htm,MS96, siegel:2002:scr}:
\begin{equation}
\rho_{\rm {halo}} \propto {\left(1 + r/a_{0,\rm {halo}}\right)^{-3.5}}
\end{equation}
\noindent where $a_{0,\rm {halo}}$ is the scale radius of 3.5 kpc. 

Once we have the mass model, we can populate the halo with our
binaries.  
We use the {\tt StarTrack} population synthesis code for single and binary evolution  
\citep{Bel08} to evolve our halo stellar population.  
We assume one metallicity and age for the entire halo, although we note 
that the halo has been observed to be comprised of two distinct components varying 
in metallicity \citep{Car07}, and may very well have a triaxial shape \citep[e.g.,][]{Helmi04}.  
For the halo population we use an evolution
model which incorporates: {\em i)} low metallicity $Z=0.0001$, {\em ii)} 
a burst of star formation at $t=0$ Gyr, and {\em iii)} is evolved through 13 Gyr
\citep[e.g.,][]{Sch06}.  Our spatial distribution and evolutionary
model parameters for the disk$+$bulge population are described in
detail in \S\, 2 of RBBLW, but we summarize the differences here: 
{\em i)} the disk and bulge stellar populations are evolved with
near-solar ($Z=0.02$) metallicity, 
{\em ii)} the disk has a constant star formation history for 10 Gyr,
the bulge has a constant star formation history for the first Gyr with
none thereafter, and {\em iii)} both disk and bulge populations are 10
Gyr old.   The remaining evolutionary parameters for disk, 
bulge and halo populations are the same.  

Once the halo stellar population has been evolved, we record
the physical properties of the close white dwarf binaries (orbital periods
$\lesssim 5.6$ hours) and calibrate the results.
We construct a grid of models in which we constrain the total mass of white dwarf stars in the
present halo relative to the total baryonic halo mass. We choose 4 different 
realizations of the halo, keeping the halo mass constant, only varying the parameter which sets 
the fraction, by mass, of white dwarf stars (single and binary) within it ($\eta_{\rm B}$). Binarity
of 50\% is assumed.  
We choose mass fractions of 0\% (no white dwarfs in the halo) and
100\% as the extreme cases. For intermediate 
cases, we choose 2 models of 15\% and 30\% (based on microlensing results of 
\citet{Alc00}\footnote{\citet{Alc00} suggest that a 20\% (by mass) white dwarf halo is consistent 
with their microlensing results, although likely it is an overestimate if compared with
chemical evolution models.}, \citet{Las00}, and \citet{Bro03}).  
We extract all double white dwarfs with GR frequencies within the {\em LISA} sensitivity range 
($10^{-4} - 0.1$ Hz) to calculate the spectral amplitudes, and compare the {\em LISA} halo spectra to 
that of the Galactic disk$+$bulge.  
We calculate the {\em LISA} timestream signals using the approach 
of \citet{RCP04}, which are added together yielding the total observatory data stream, which is 
then Fourier transformed in order to produce the frequency domain data (the spectra presented in 
\S~\ref{results}). 
The actual {\em LISA} GR spectra is obtained using the \citet{BDL04} simulation code.  
We note that all of our white dwarf binaries within the {\em LISA} sensitivity range have 
circular orbits. Eccentric white dwarf binaries are expected to arise from dynamical interactions 
in globular clusters \citep{Ben01}, where the phase space densities are much higher than in 
the halo or the disk. These eccentric binaries, however, could provide a unique opportunity 
for learning about white dwarf structure with {\em LISA} \citep{Wil07}.  

\section{RESULTS}
\label{results}

For $\eta_{\rm B}=1$ in a $10^9 \msun$ halo, we obtain $1.5 \times 10^{9}$ white dwarfs (single and binary); 
$500 \times 10^6$ binary white dwarfs and out of these $27.5 \times 10^6$ {\em LISA} binary white dwarfs.
Only $5.5\%$ of double white dwarfs have periods shorter than 5.6 hr.    
Obviously, the above numbers scale down linearly with $\eta_{\rm B}$.  
Chemo-dynamical simulations of the Milky Way \citep{Bro03} have demonstrated that a white dwarf 
dominated halo that is evolved from a white dwarf progenitor-dominated initial mass function \citep{CSM96} 
at early times would lead to an overproduction of carbon and nitrogen when compared to 
observed abundances, so the $\eta_{\rm B}=1$ model is unrealistic.  For a realistic
upper limit on the white dwarf halo contribution we choose $\eta_{\rm B}=0.3$ based on \citet[][their figure 2; also 
Brook et al. (2003) their \S\,5]{Las00}, and this yields $8.3 \times
10^{6}$ {\em LISA} double white dwarfs in the halo.  For a Galactic
disk$+$bulge with a total stellar mass of $6 \times 10^{10} \msun$ for
{\em all} stellar types \citep{klypin:2002:mwmass}, we predict a total 
of $\sim 1.6 \times 10^{9}$ white dwarfs; $550 \times 10^{6}$ double
white dwarfs, out of which only $8\%$ are found
within the {\em LISA} band: $44.5 \times 10^{6}$ (see RBBLW \S\, 3).

Note that the halo is presumed to include a specific fraction
of mass in white dwarfs ($\eta_{\rm B}$),
therefore the predicted numbers of white dwarfs are a direct result of the {\em i)} adopted halo mass
and {\em ii)} calculated (with population synthesis) mass and period distributions for
halo white dwarf binaries with the assumed binary fraction.  For the
rest of the Galaxy, in addition to the calculation of double white dwarf properties, 
we have computed the white dwarf mass fraction with the adopted Galactic field 
initial mass function (for details see RBBLW).  In other words the
disk$+$bulge model results in a true white dwarf formation efficiency 
per unit mass, while in the halo model this efficiency is imposed 
a priori (through the straight forward application of observational 
constraints, e.g., MACHOs).

The disk and halo double 
white dwarf populations differ significantly in numbers and physical properties, due to the different environments 
under which stellar evolution proceeds (metallicity, age, star formation history).  For example, 
typical average double white dwarf 
chirp masses (${\cal M} = \left(M_{\rm p}M_{\rm s}\right)^{3/5}/\left(M_{\rm p}+M_{\rm s}\right)^{1/5}$, 
where $M_{\rm p}$ and $M_{\rm s}$ represent the first formed and second formed white dwarf masses, respectively)
of halo systems are $0.13 \msun$ as compared to $ 0.19 \msun$ 
for the disk.  Also, there are relatively few short period 
double white dwarfs in the halo (5.5\% vs. 8.0\% for the disk$+$bulge) since this population is older 
and a larger number of short period systems have merged.
In particular, some double white dwarf systems form on rather short
orbits (e.g., hybrid white dwarfs\footnote{Carbon-oxygen (CO) white dwarfs with a thick helium
envelope.} with carbon-oxygen white dwarf companions),  and none of these systems   
are found in our 13 Gyr old halo population (see below).  

In Figure~\ref{dndf} we show the number density (per resolvable
frequency bin) of {\em LISA} white dwarf binaries as a function of GR 
frequency for both the halo and the combined disk$+$bulge population
of RBBLW.  At nearly all frequencies, the disk$+$bulge
population outnumbers the halo by nearly a factor of two ($\sim$ 45
vs. $\sim$ 28 million {\em LISA} binaries).  However, there is a
relative increase in the number of halo systems between $\sim 0.0002 -
0.0004$ Hz ($\sim 170 - 80$ minute orbital periods).  This is
attributed to the fact that there is a relatively higher number of 
RLOF double white dwarfs with hydrogen white dwarf donors in the halo
population.  Double degenerate binaries with hydrogen white dwarfs
take a long time to form; on the order of $\gtrsim 10^{9}-10^{10}$ 
years as opposed to $\sim 10^{8}$ for other, heavier double WD 
types (e.g., CO-CO, hybrid-CO) descended from more massive progenitors.  
Remnant binaries formed from progenitors with more massive stars 
are more common in the (younger) Galactic disk.  Hydrogen 
white dwarfs are evolved from binary progenitors in which a 
white dwarf (e.g., carbon-oxygen or helium) is
feeding from a low-mass main sequence star.  At some stage during the mass 
transfer,\footnote{Mass accretion rates are $\sim 10^{-11}$ M$_{\odot}$
  yr$^{-1}$.}
the main sequence donor becomes depleted of enough mass
- to a mass below that of the hydrogen-burning limit -  
such that it is no longer capable of fusing hydrogen in
its core and thus becomes degenerate (a hydrogen white dwarf is born;
specific details about the evolution of these systems  
can be found in \S\, 3.1 of RBBLW).  
Because all of the binaries in
the halo are 13 Gyr old, most pre-stable RLOF systems have had time to
reach contact, and many binaries which have evolved from more
massive progenitors have since merged.  Double white dwarfs involving
hydrogen donors make up 75 \% of our {\em LISA} binary white dwarfs in the 13 Gyr 
old halo, spanning a frequency range from $\sim 0.0002 - 0.0009$ Hz.  
By contrast, white dwarf binaries with hydrogen white
dwarfs only make up 40 \% of the slightly younger bulge population 
($9 - 10$ Gyr) and only 13 \% of the younger disk population.
Additionally, only the disk population contains close white dwarf
binaries with $f_{\rm gr} > 4.4$ mHz (orbital periods less than 8
minutes), the majority of which are CO white dwarfs accreting from
helium white dwarfs (a class of AM CVn binaries; \citet{War95}).  
Regarding massive white dwarfs, only 1.5 \% of {\em LISA} halo 
double white dwarfs are CO+CO systems, where as this fraction 
is 12 \% for the disk.  Out of the halo CO+CO systems, $\sim 2/3$ will
merge within a Hubble time and have combined masses above $1.4$
M$_{\odot}$, making them potential double degenerate scenario 
Type Ia supernovae \citep{IT84,Web84}.  

To obtain a better idea of the effect that location in the Milky Way 
has on the GR amplitude, in Figure~\ref{dist} we show the
number density distribution as a function of distance from the Sun for 
the three halo realizations, as well as the disk and bulge populations
of RBBLW.  It is immediately obvious that the halo signal is 
expected to be less than that of the rest of the Milky Way, given the 
extensive yet sparse stellar density distribution with respect to the bulge
and disk.  It becomes
clear that even though the number of halo double white dwarfs are only
about a factor of $1.6$ in number below the disk$+$bulge combined, the
GR amplitude is expected to be much weaker given the $1/D$ dependence
\citep{RCP04}, and the fact that the physical properties of the double
white dwarfs (chirp masses, separations) in all three populations are not
drastically different.

In Figure~\ref{sig} we show spectral amplitude vs. GR frequency 
of three Galactic halo double white dwarf realizations: 15\%, 30\%  and 100\% 
halo models. In addition to our halo signals we show the {\em LISA} sensitivity 
curve,\footnote{\emph{Online Sensitivity Curve Generator}, based on Larson, Hiscock \& Hellings, 
\texttt{http://www.srl.caltech.edu/$\sim$shane/sensitivity/}}
the median signal arising from the Galactic population from RBBLW (see
their section 3.4.1),
and the signal of the double white dwarf 
foreground of \citet{Nel04}. The shape of the {\em LISA} instrumental noise curve 
is a function of the acceleration noise (from {\em LISA}'s accelerometers), position 
noise and the gravitational wave transfer function (see \citet{LHH00} for a more detailed 
description of how the curve is calculated).  The 
signal presented in \citet{Nel04} is a measure of the barycentred double white dwarf {\em confusion foreground} 
amplitude $h$ (rather than spectral amplitude $h_{\rm f}$, see \citet{Tim06}), and is artificially truncated 
beyond $\sim 2$ mHz where individual binaries become resolved and the signal 
is no longer confusion-limited in that study, where as the signal from RBBLW is shown for a range 
of frequencies.  We have scaled the amplitude of \citet{Nel04} by 
$\sqrt{T_{\rm obs}} \times \sqrt{3/20}$, accounting for a 1-yr observation time for {\em LISA} 
and signal modulation due to the motion of {\em LISA}, respectively, to arrive at the root spectral density 
(spectral amplitude) $h_{\rm f}$.  We note that for even our most 
extreme (and unphysical) halo realization of $\eta_{\rm B}=1$, the halo signal is 
well below the signals of both \citet{Nel04} and RBBLW, as well as the
{\em LISA} sensitivity curve.

\section{DISCUSSION}
\label{discussion}

The amplitude of the GR signal from halo white dwarfs was previously estimated
by \citet{His00}.  In that study, \citet{His00}
used \citet{Hil90} to estimate the number of disk white dwarfs, 
and arrived at the number (both single and binary) $N_{\rm disk}=6.5 \times 10^{8}$.
\citet{His00} assumed that the properties of the halo white dwarfs were
the same as those of the disk. Next they adopted the number of halo white dwarfs to
be $N_{\rm halo}=2 \times 10^{11}$. This was based on the mass of MACHOs
($2 \times 10^{11}$ halo objects with masses $\sim 0.1-1 \msun$ derived by the
MACHO collaboration \citep{Alc00}), and the assumption that {\em all} MACHOs are white dwarfs.
Note that this results in a halo mass of $\sim 10^{11} \msun$;
\citet{His00} does not present the actual numbers of double white dwarfs that are within 
the {\em LISA} band for the halo nor for the disk populations.
Naturally, \citet{His00} obtain, due to a very high number of halo white dwarfs,
a very strong GR signal from the halo population. In fact they estimated
that the level of signal from the halo could be a factor of $\sim 5$ stronger
than the one arising from the disk.

Comparison of predictions clearly shows that our number for the {\em entire} 
disk$+$bulge white dwarf population ($\sim 1.6 \times 10^9$) is similar to  
the disk prediction of \citet{His00}. However, their number for the entire halo white dwarf population is  
$\sim 2$ orders of magnitude higher than we can presently use based on star count data 
\citep{zinn:1985:gcs,saha:1985:rls,morrison:1996:htm,MS96, siegel:2002:scr}: 
Even if we place the entire baryonic halo mass of $10^9 \msun$ in white dwarfs we
obtain only $1.5 \times 10^9$ white dwarfs in the halo, while for a more realistic 
halo model ($\eta_{\rm B} \sim 0.15$) our number is $2.25 \times 10^8$ (see
\S\,3).
Such a large discrepancy in the estimate of the number of white dwarfs in the halo leads to a very
different final result. In particular, our halo contribution is never
higher than the Galactic disk$+$bulge signal. This marked difference stems from the fact that we
employ a much smaller (baryonic) mass of the halo ($10^{9} \msun$) than
\citet{His00} ($10^{11} \msun$).  

The simple fact is that with a Milky Way virial mass of approximately $10^{12}$ M$_\odot$ 
\citep{klypin:2002:mwmass,li:2007:mwmass}, the maximum baryonic mass in the {\it entire} Galaxy cannot 
be more than $\sim 8 \times 10^{10}$ M$_\odot$ without violating the strong constraint on $\Omega_B$ set by 
WMAP3 \citep{Spe07}. Given that the Milky Way disk and bulge itself is known to have a mass $\sim 6 
\times 10^{10}$ M$_\odot$ \citep{klypin:2002:mwmass}, it is extremely
unlikely that the baryonic halo can be as massive as $O(10^{11})$ M$_\odot$. 
Even if there were significant play in the total baryonic mass of the halo, a white dwarf population of $O(10^{11})$ M$_\odot$ 
would have had to have lost approximately $10^{10}$ M$_\odot$ in gas during
the planetary nebulae phase, and this much gas is likely to have been observed. 

We have calculated the {\em LISA} gravitational radiation signal predicted to arise                                       
from double white dwarfs in the Galactic halo.  In doing so, we have performed the                                         
first detailed calculation of the halo double white dwarf population and compared its                                        
signal to that of the disk$+$bulge population.  Thus, for the first time there is a
complete model Milky Way Galaxy (disk, bulge, halo) calculated
self-consistently with the same binary evolution population synthesis code.    
The evolutionary calculations were                                           
done with the population synthesis code {\tt StarTrack} \citep{Bel08}, and                                        
the GR signal calculations were obtained with the detailed {\em LISA} simulation code of 
\citet{BDL04}. It was found that the GR                                         
signal arising from the halo population is significantly smaller                                                    
than that of the rest of the Galaxy, and will not contribute
substantially to
the overall Galactic foreground signal.  Further, if we use recent microlensing                                           
results in order to constrain the mass of the halo white dwarf population to                                        
$30\%$ of the halo baryonic mass, we predict that the GR signal                                                
arising from the halo is at the level of $h_{\rm f} \approx 7.1 \times 10^{-21}$                                         
Hz$^{1/2}$ at 1 mHz (see Figure~\ref{sig}). The disk$+$bulge double white dwarf population generates 
a much larger noise level: $h_{\rm f} \sim 10^{-19}$ Hz$^{1/2}$ at                                        
1 mHz both for the {\tt StarTrack} disk$+$bulge population (Ruiter et al. 2007) as well as        
for the previous prediction for the combined Galactic disk$+$bulge
population obtained with a different population synthesis model                                      
\citep{Nel04}. 

\begin{figure*}
\includegraphics[width=\textwidth]{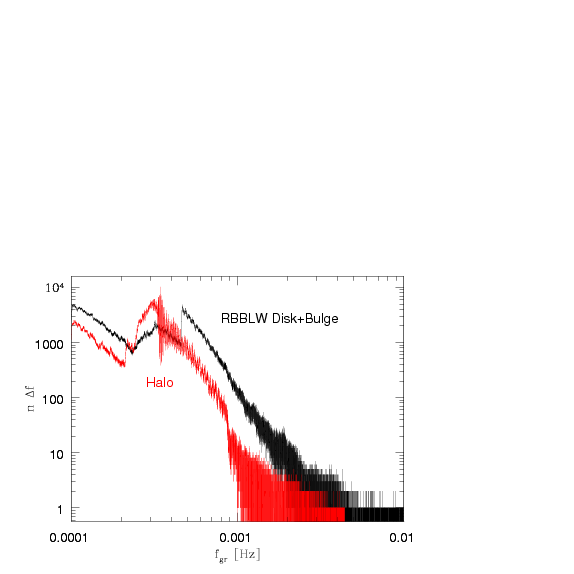}
\caption{Number density ($n = (dN/df); \delta \, f $ is the size of a
  resolvable frequency bin for a 1 year observation time, $1/T_{obs} =
  30$ nHz) for the 100 \% model halo population of {\em LISA} double white dwarfs,
  and the entire Milky Way disc+bulge population of {\em LISA} double white
  dwarfs presented in Ruiter et al. (2007).  On average, the
  disc+bulge population is a factor of $\sim 2$ greater in number than the 100 \% halo population.}
\label{dndf}
\end{figure*}

\begin{figure*}
\includegraphics[width=\textwidth]{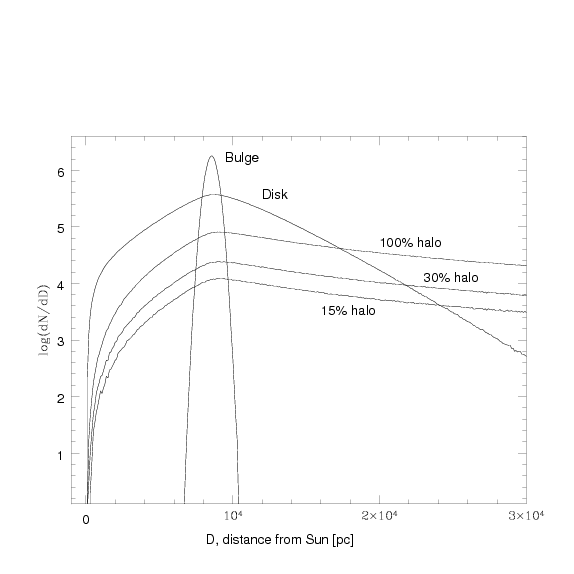}
\caption{Number of {\em LISA} white dwarf binaries as a function of 
  distance from the Sun in bin sizes of 100 pc.  The Galactic bulge
  distance is concentrated around 8.5 kpc, and there are no more disc
  systems beyond $\sim $ 50 kpc.  The 100\% halo realization 
  is nearly an order of magnitude below that of the disc distribution
  for the potentially strongest (closest) GR sources, until $> 10$ kpc, 
  where the disc begins to fall off more steeply.}
\label{dist}
\end{figure*}

\begin{figure*}
\includegraphics[width=\textwidth]{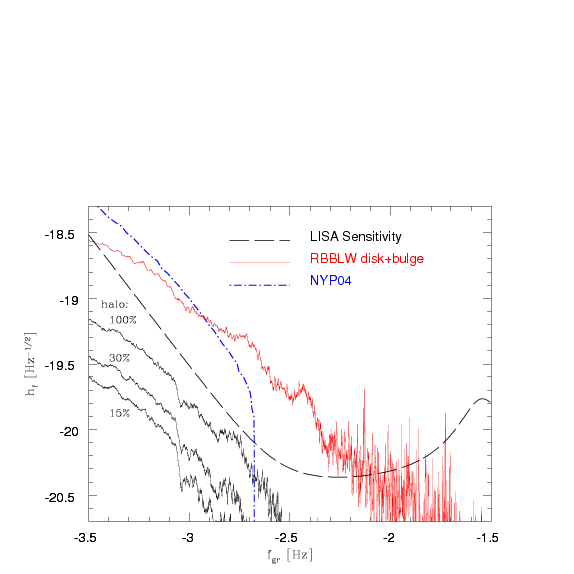}
\caption{{\em LISA} gravitational wave spectra (amplitude densities). Shown are the gravitational 
  wave signal for 15\%, 30\% and 100\% (i.e., $\eta_{\rm B}=$ 0.15,
  0.3 and 1, smoothed over 5000 resolvable frequency bins) 
  realizations of the halo population of {\em LISA} double white dwarfs computed from our population synthesis models;  
  Galactic double white dwarf signal from Ruiter et al. 2007 (red); smoothed
  Galactic disc foreground from Nelemans et al. 
  (2004) truncated beyond $\sim 2$ mHz where sources start to become
  resolved (blue dot-dash); and the {\em LISA} sensitivity curve for a signal
  to noise ratio of 1 (dashed line).}
\label{sig}
\end{figure*}

Therefore, throughout the low-frequency region where the 
disk$+$bulge signal is confusion-limited, the halo signal is a factor
of $\sim 10$ (1.1 in log compared to the foreground of Ruiter et
al. (2007)) lower than that of the disk$+$bulge. Since we have used 
an upper limit on the white dwarf contribution in the halo, and the actual 
halo white dwarf content is probably not higher than $\sim 10-20\%$, (\citet{Tis07}; see also \citet{Tor02} for an estimate 
and discussion of the number density of halo white dwarfs), 
we predict that the actual halo signal will be more than $10 \times$                                            
lower than that of the disk and bulge combined. The reduced number of 
high-frequency halo systems compared with the 
disk population will result in a small number of potentially resolvable 
systems from the halo, 
since no halo systems are found with GR frequencies above 4.4
mHz (see Figure~\ref{dndf}).  A number of Galactic binaries predicted to be
resolved with {\em LISA} 
have frequencies above this value (log($f$)$\approx -2.35$; \citet{Nel04}).   
Even for an unrealistic halo model for which $\eta_{\rm B}=1$, the level of                                            
the average halo GR signal does not significantly approach that of the
disk$+$bulge, and remains below the {\em LISA} sensitivity curve.    
It is clear that {\em LISA}'s ability to detect other sources will not
be strongly curtailed by halo double white dwarfs, and that the GR signal from 
white dwarf binaries in the rest of the Galaxy (primarily the disk)
will still constitute the prime limiting confusion foreground for {\em LISA}.                                

\acknowledgments
AJR is thankful to S. Torres for informative discussion.  MJB acknowledges the support of NASA 
through grant NNG94GD52G  and the Center for Gravitational Wave Astronomy (NAG5-13396). We also 
thank S. Larson for {\em LISA} sensitivity curve data, and G. Nelemans for the use of Nelemans et al. (2004) 
{\em LISA} amplitude data and helpful discussion.  The {\tt StarTrack}
simulations were run at the Nicolaus Copernicus Astronomical Center
and the Advanced Center for Computation, Research and Education.

\clearpage

\end{document}